\documentclass[aps,plb, nofootinbib, notitlepage,onecolumn,superscriptaddress]{revtex4-1}
\usepackage{bm}
\usepackage{latexsym}
\usepackage{amsmath,amsfonts,amssymb}
\usepackage{graphicx,epsfig}
\usepackage{psfrag}
\usepackage{amsthm}
\usepackage{color}
\usepackage{braket}
\usepackage[normalem]{ulem}
\usepackage{hyperref}
\usepackage[titletoc,title]{appendix}
 \usepackage{environ}
\usepackage{mathtools}
\usepackage{float}
\usepackage[makeroom]{cancel}
\usepackage{bbold}
\usepackage{footmisc}
\usepackage{titlesec}

\usepackage[utf8]{inputenc}
\usepackage{lineno}
\newcommand{\be}{\begin{equation}}
\newcommand{\ee}{\end{equation}}
\newcommand{\bea}{\begin{eqnarray}}
\newcommand{\eea}{\end{eqnarray}}

\makeatletter
\newcommand\setcurrentname[1]{\def\@currentlabelname{#1}}
\makeatother

\NewEnviron{eqn}{
\begin{align}
\begin{split}
  \BODY
\end{split}
\end{align}
}

\NewEnviron{eqn*}{
\begin{align*}
\begin{split}
  \BODY
\end{split}
\end{align*}
}

\newcommand{\tilbox}{%
  \mspace{2mu}%
  \widetilde{\mspace{0mu}\rule{0pt}{1.4ex}\smash[t]{\Box}}%
}

\begin{document}
%%%%%%%%%%%%%%%%%%%%%%%%%%%%%%%%%%%%%%
\title{Testing the Quantum Equivalence Principle with Gravitational Waves}
%
%\title{The generalized Quantum Equivalence Principle}
%%%%%%%%%%%%%%%%%%%%%%%%%%%%%

\author{Saurya Das} 
\email{saurya.das@uleth.ca}
\affiliation{Theoretical Physics Group and Quantum Alberta, Department of Physics and Astronomy,
University of Lethbridge,
4401 University Drive, Lethbridge,
Alberta, T1K 3M4, Canada}

\author{Mitja Fridman}
\email{fridmanm@uleth.ca}
\affiliation{Theoretical Physics Group and Quantum Alberta, Department of Physics and Astronomy,
University of Lethbridge,
4401 University Drive, Lethbridge,
Alberta, T1K 3M4, Canada}

\author{Gaetano Lambiase} \email{lambiase@sa.infn.it}
\affiliation{Dipartimento di Fisica E.R: Caianiello, Universit\'a di Salerno, Via Giovanni Paolo II, 132 - 84084 Fisciano, Salerno, Italy}
\affiliation{INFN - Gruppo Collegato di Salerno, Via Giovanni Paolo II, 132 - 84084 Fisciano, Salerno, Italy}

\begin{abstract}
\begin{center}
    \textbf{Abstract}\\
\end{center}
\par\noindent
We study modifications of gravitational wave observables, such as the wave amplitude and frequency, which follow from the quantum equivalence principle, and
are expressed in terms of the inertial, gravitational and rest masses of the LIGO/Virgo mirrors. We provide bounds on the violations of the quantum equivalence
principle by comparing the results with the most resolved gravitational wave events observed by the LIGO/Virgo collaboration. The formalism is equally applicable
to other future ground and space-based gravitational wave detectors.
\end{abstract}
\maketitle
%
%%%%%%%%%%%%%%%%%%%%%%%%%%%%%%%%%%%%%%%%%%%%%%%%%%%%%%
\section{Introduction}

The problem of Quantum Gravity (QG) is considered to be one of the most important in theoretical physics. 
While quantum theory and General Relativity (GR) are two of the most successful theories of Nature, explaining and predicting phenomena in their respective domains with high accuracy \cite{Zeilinger:1999zz,CMW}, it is not an easy task to combine the two into a single consistent theory of QG. Such a theory of QG is required to explain the physics at the centers of black holes, as well that of the early Universe and other high-energy phenomena. Several candidate theories of QG, among which String Theory \cite{qg1} and Loop Quantum Gravity \cite{Smolin:2004sx} are some of the most studied, attempt to address the QG problem. However, they can not be tested directly, due to the required high energies (around the Planck energy $E_P\sim10^{16}\,\mathrm{TeV}$) for the QG effects to manifest, which are out of reach in currently available experiments. Until such energies can be reached, we can search for QG signatures with phenomenological approaches, or formulating fundamental principles of QG. One such principle is the Quantum Equivalence Principle (QEP). 
%Such an approach is conceptual, where a foundational principle is formulated, which should form the basis of a consistent theory of QG.
The classical equivalence principle is the foundation based on which GR is formulated \cite{CMW,Weinberg:1972kfs,Einstein}. In the same way it can be argued that 
%, as we shall show in this article, 
the QEP as stated below
can be considered as 
%the basis for construct 
a foundational principle for formulating QG itself.

%The classical Equivalence Principle is the foundational principle on which GR is formulated \cite{CMW,Weinberg:1972kfs,Einstein}.
A quantum formulation of the equivalence principle is the so-called QEP. One realization of the QEP has been proposed by Zych \& Brukner \cite{Zych:2015fka}, where the mass of a test particle in the theory is 
associated with three distinct masses,
%distinguished in three different masses, 
%which appear in different contexts. 
namely the inertial mass $m_I$, the gravitational mass $m_G$ and the rest mass $m_R$. These masses are then promoted to quantum operators as \cite{Zych:2015fka}
\begin{eqnarray}
\label{massop}
m_\alpha\longrightarrow\hat{M}_\alpha=m_\alpha\hat{\mathbb{1}}+\frac{\hat{H}_{\mathrm{int,\alpha}}}{c^2}~,
\end{eqnarray}
where $\alpha=\mathrm{I,G,R}$ denote the inertial, gravitational and rest masses of the test particle, respectively,
and $c$ the speed of light. The $m_\alpha$ are interpreted as the ground state masses, and $\hat{H}_{\mathrm{int,\alpha}}$ describe the unknown internal energy operators of the composite test particle. The statement of the QEP is the equality of these mass operators \cite{Zych:2015fka}
\begin{eqnarray}
\hat{M}_\mathrm{I}=\hat{M}_\mathrm{G}=\hat{M}_\mathrm{R}~.
\label{qep10}
\end{eqnarray}
The above equalities encompass the quantum versions of
the three aspects of the Equivalence Principle. Namely, $\hat{M}_\mathrm{I}=\hat{M}_\mathrm{G}$ is the quantum version of the weak equivalence principle (WEP),
$\hat{H}_{\mathrm{int,R}}=\hat{H}_{\mathrm{int,I}}$ rerpesents the quantum version of the local Lorentz invariance (LLI) and $\hat{H}_{\mathrm{int,R}}=\hat{H}_{\mathrm{int,G}}$ the quantum version of the local position invariance (LPI). Note that Ref. \cite{Zych:2015fka} applies the above QEP formulation to a non-relativistic particle in a weak gravitational field. The generalization of the QEP formalism to relativistic speeds, strong gravitational fields and different particle statistics was done in Ref. \cite{Das:2023cfu}. 

Note, however, that to establish the above formulation of the QEP as the foundational principle of QG, it must be verified 
experimentally in as many physical scenarios as possible. 
One such realm in which the 
%of the most interesting types of experiments, where the 
QEP can be tested is in gravitational wave (GW) detectors, since they are extremely precise and in fact measure the effects of gravity, which is an essential ingredient of the QEP. 

GWs have been used extensively to test fundamental physics and modified theories of gravity \cite{Feng:2016tyt,Das:2021lrb,Bhattacharyya:2020ooz,Das:2022hjp,Moussa:2021qlz, Gogoi:2020,Katsuragawa:2019uto,Kalita:2021zjg,Lambiase:2020vul,Tino:2020nla}. However, the above approaches to modify the GW formalism are not directly applicable to the QEP, because they do not generally include the properties of the test particle, i.e., the LIGO/Virgo detector mirrors, which as shown
in Ref. \cite{Das:2023cfu}, are 
%are 
the key components in testing the QEP. 
In this work we approach testing the QEP using GW physics. We apply the general QEP formalism to the mathematical framework of GWs and consider the expectation values of $\hat{M}_\alpha$ for phenomenological predictions, which in principle test the classical Einstein Equivalence Principle (EEP).
%, as obtained from GR. 
The EEP violations are then bounded by considering the most resolved GW events, measured by the LIGO/Virgo collaboration. The bounds on 
%the quantum versions of 
WEP, LLI and LPI turn out to be of the order $\sim10^{-1}$, which is due to the high uncertainty in the measurements of the source parameters. Although, the above bound may not seem as strong as one would hope, the experimental precision of the GW detectors will continue to increase with time, and the formalism 
developed here will apply equally well to future ground and space-based GW detectors. Note that throughout this work, the metric signature convention of $(+,-,-,-)$ is used.

This work is structured as follows. In Section \ref{sec:mods} we introduce the QEP modifications to the GW formalism, while we consider the case for inspiraling binaries in Section \ref{sec:binaries}. In Section \ref{sec:bounds} we obtain bounds on the classical EEP violations from observed GW events, and in Section \ref{sec:conc} we summarize with concluding remarks.
%%%%%%%%%%%%%%%%%%%%%%%%%%%%%%%%%%%%%%%%%%%%%%%%%%%%%%
\section{Modifications of the linearized Einstein equations}
\label{sec:mods}

In order to apply the QEP formalism in the context of GWs, we show how any violation of the QEP modifies the underlying GW theory and the related observables. This work uses the procedure adopted in 
Ref. \cite{Maggiore:2007ulw}.
%follows the derivation of the above mentioned theory from Ref. \cite{Maggiore:2007ulw}. 
The QEP modifications of GR were derived in the context of the general formalism of the QEP in Ref. \cite{Das:2023cfu}. Note that \textit{QEP modifications} refer to mass ratio corrections before mass quantization is explicitly considered. The QEP modified Einstein equations read as \cite{Das:2023cfu}
\begin{eqnarray}
\label{qepeinstein}
R_{\mu\nu}-\frac{1}{2}R\,g_{\mu\nu}=\frac{m_\mathrm{G}m_\mathrm{I}}{m_\mathrm{R}^2}\,\frac{8\pi G}{c^4}\,T_{\mu\nu}~.
\end{eqnarray}
%\textcolor{red}{\bf GL: 1) Maybe here we can explain why the factor $\frac{m_\mathrm{G}m_\mathrm{I}}{m_\mathrm{R}^2}$ appears. 2) How the masses $m_G, m_I. m_R$ are interepreted (or equivalently, ae they the mass of what)?}
From the above, we obtain the QEP modified linearized Einstein equations, for which $g_{\mu\nu}=\eta_{\mu\nu}+h_{\mu\nu}$ ($\eta_{\mu\nu}$ is the Minkowski metric and $h_{\mu\nu}$ the linear perturbation with $|h_{\mu\nu}|\ll1$) in the Lorentz gauge $\partial^\nu \bar{h}_{\mu\nu}=0$ as
\begin{eqnarray}
\label{lineinst}
    \tilbox\,\Bar{h}_{\mu\nu}=-\frac{m_Gm_I}{m_R^2}\,\frac{16\pi G}{c^4}\,T_{\mu\nu}~,
\end{eqnarray}
where $\tilbox=\tilde{\partial}_\mu\tilde{\partial}^\mu=\frac{1}{c^2}\partial_0^2-\frac{m_R}{m_I}\boldsymbol{\partial}^2$ \cite{Das:2023cfu} %\textcolor{red}{\bf GL: Also here we can explain why $\frac{m_R}{m_I}$ appears - A question that arises: does such a factor have some influence in other physics framework?} 
is the modified D'Alembert operator and $\Bar{h}_{\mu\nu}=h_{\mu\nu}-\frac{1}{2}\eta_{\mu\nu}\,h$ with $h=\eta^{\mu\nu}h_{\mu\nu}$. To obtain the solution for the above second-order partial differential equation, one usually uses Green's function approach. It turns out that Green's function will obtain QEP modifications due to the modified D'Alembert operator. The definition of the modified Green's function then reads as
\begin{eqnarray}
\label{greendef}
    \tilbox\,G(x-x')=\delta^{(4)}(x-x')~,
\end{eqnarray}
which is used to solve Eq. (\ref{lineinst}) as
\begin{eqnarray}
    \Bar{h}_{\mu\nu}(x)=-\frac{m_Gm_I}{m_R^2}\,\frac{16\pi G}{c^4}\int \mathrm{d}^4x'\,G(x-x')\,T_{\mu\nu}(x')~.
\end{eqnarray}
By using the standard approach to solve Eq. (\ref{greendef}), it turns out that the Green's function $G(x-x')$ obtains modifications, due to the modified $\tilbox$ operator as
\begin{eqnarray}
    G(x-x')=\frac{m_I}{m_R}\,\frac{1}{4\pi}\,\frac{1}{|\mathbf{x}-\mathbf{x}'|}\,\,\delta\!\left(\sqrt{\frac{m_R}{m_I}}\,ct-|\mathbf{x}-\mathbf{x}'|-\sqrt{\frac{m_R}{m_I}}\,ct'\right)~.
\end{eqnarray}
It is standard practice to simplify the theory further using the so-called transverse-traceless (TT) gauge, where $\bar{h}^{0\mu}=0$, $\Bar{h}_i^i=0$ and $\partial^j\Bar{h}_{ij}=0$, which is valid only outside the source \cite{Maggiore:2007ulw}. The modified $\Bar{h}_{\mu\nu}$ can then be written in the TT gauge as %
\begin{eqnarray}
\label{htt}
    h_{ij}^{TT}(t,\mathbf{x})=-\frac{m_Gm_I^{2}}{m_R^3}\,\frac{4G}{c^4}\,\frac{1}{r}\,\Lambda_{ij,kl}(\hat{\mathbf{n}})\int\mathrm{d}^3x\,\,T_{kl}\!\left(t-\sqrt{\frac{m_I}{m_R}}\,\frac{r}{c}+\sqrt{\frac{m_I}{m_R}}\,\frac{\hat{\mathbf{n}}\cdot\mathbf{x}'}{c},\mathbf{x}'\right)~,
\end{eqnarray}
where
\begin{eqnarray}
    \Lambda_{ij,kl}(\hat{\mathbf{n}})=\delta_{ik}\delta_{jl}-\frac{1}{2}\delta_{ij}\delta_{kl}-n_jn_l\delta_{ik}-n_in_k\delta_{jl}+\frac{1}{2}n_kn_l\delta_{ij}+\frac{1}{2}n_in_j\delta_{kl}+\frac{1}{2}n_in_jn_kn_l~,
\end{eqnarray}
which relates $\Bar{h}_{\mu\nu}$ to the TT gauge as $h_{ij}^{TT}=\Lambda_{ij,kl}\Bar{h}_{kl}$ and $\hat{\mathbf{n}}$ is the direction of propagation. It is convenient to write Eq. (\ref{htt}) in terms of the frequency distribution of the source, which is achieved by considering the Fourier transform of the spatial component of the energy-momentum tensor $T_{kl}$, and reads as
\begin{eqnarray}
\label{httft}
    h_{ij}^{TT}(t,\mathbf{x})=-\frac{m_Gm_I^{5/2}}{m_R^{7/2}}\,\frac{4G}{c^5}\,\frac{1}{r}\,\Lambda_{ij,kl}(\hat{\mathbf{n}})\int_{-\infty}^\infty\frac{\mathrm{d}\omega}{2\pi}\,\,\tilde{T}_{kl}\!\left(\omega,\sqrt{\frac{m_I}{m_R}}\,\frac{\omega}{c}\,\hat{\mathbf{n}}\right)e^{-i\omega\left(t-\sqrt{{m_I}/{m_R}}\,r/c\right)}~.
\end{eqnarray}
The above is used to compute the radiated energy in a solid angle $\mathrm{d}\Omega$ relative to the source as (see Appendix \ref{app:qepenspec})
\begin{eqnarray}
\label{genenmod}
    \frac{\mathrm{d}E}{\mathrm{d}\Omega}&=&\frac{m_R^{3/2}}{m_Gm_I^{1/2}}\,\frac{c^3r^2}{32\pi G}\int_{-\infty}^\infty\mathrm{d}t\,\Dot{h}_{ij}^{TT}\Dot{h}_{ij}^{TT}%= \frac{m_R^{3/2}}{m_Gm_I^{1/2}}\frac{r^2c^3}{16\pi G}\int_{-\infty}^\infty\mathrm{d}t\,\left(\Dot{h}_{+}^{2}+\Dot{h}_{\times}^{2}\right) 
    \nonumber \\
    &=&\frac{m_Gm_I^{9/2}}{m_R^{11/2}}\,\frac{G}{2\pi^2 c^7}\,\Lambda_{ij,kl}(\hat{\mathbf{n}})\int_0^\infty\mathrm{d}\omega\,\omega^2\,\,\tilde{T}_{ij}\!\left(\omega,\sqrt{\frac{m_I}{m_R}}\,\frac{\omega}{c}\,\hat{\mathbf{n}}\right)\,\tilde{T}_{kl}^*\!\left(\omega,\sqrt{\frac{m_I}{m_R}}\,\frac{\omega}{c}\,\hat{\mathbf{n}}\right)~,
\end{eqnarray}
which can be rearranged to obtain the energy spectrum over frequencies as
\begin{eqnarray}
    \frac{\mathrm{d}E}{\mathrm{d}\omega}=\frac{m_Gm_I^{9/2}}{m_R^{11/2}}\,\frac{G\omega^2}{2\pi^2c^7}\int\mathrm{d}\Omega\,\Lambda_{ij,kl}(\hat{\mathbf{n}})\,\,\tilde{T}_{ij}\!\left(\omega,\sqrt{\frac{m_I}{m_R}}\,\frac{\omega}{c}\,\hat{\mathbf{n}}\right)\,\tilde{T}_{kl}^*\!\left(\omega,\sqrt{\frac{m_I}{m_R}}\,\frac{\omega}{c}\,\hat{\mathbf{n}}\right)~.
\end{eqnarray}
Due to the fact that the energy-momentum tensor is non-vanishing only inside the source, it turns out that the dominant contribution of Eq. (\ref{htt}) arises from frequencies which satisfy $\sqrt{\frac{m_I}{m_R}}\frac{\omega}{c}\,\mathbf{x}'\cdot\hat{\mathbf{n}}\ll1$ \cite{Maggiore:2007ulw}, which is used to expand Eq. (\ref{htt}) as
\begin{eqnarray}
\label{httsmom}
    h_{ij}^{TT}(t,\mathbf{x})=-\frac{m_Gm_I^{2}}{m_R^3}\,\frac{4G}{c^4}\,\frac{1}{r}\,\Lambda_{ij,kl}(\hat{\mathbf{n}})\left(S^{kl}(t_{ret})+\sqrt{\frac{m_I}{m_R}}\,\frac{1}{c}\,n_m\,\Dot{S}^{kl,m}(t_{ret})+\frac{m_I}{m_R}\,\frac{1}{2c^2}\,n_mn_p\,\Ddot{S}^{kl,mp}(t_{ret})+\cdots\right)~,
\end{eqnarray}
where $t_{ret}=t-\sqrt{\frac{m_I}{m_R}}\,\frac{r}{c}$ is the so-called retarded time and
\begin{eqnarray}
    S^{ij}(t)&=&\int\mathrm{d}^3x\,T^{ij}\!(t,\mathbf{x}) \\
    S^{ij,k}(t)&=&\int\mathrm{d}^3x\,T^{ij}\!(t,\mathbf{x})\,x^k \\
    S^{ij,kl}(t)&=&\int\mathrm{d}^3x\,T^{ij}\!(t,\mathbf{x})\,x^kx^l~.
\end{eqnarray}
We can also define the related QEP modified moments as
\begin{eqnarray}
\label{m}
    M(t)&=&\frac{m_I}{m_R}\,\frac{1}{c^2}\int\mathrm{d}^3x\,T^{00}(t,\mathbf{x}) \\
    \label{mi}
    M^i(t)&=&\frac{m_I}{m_R}\,\frac{1}{c^2}\int\mathrm{d}^3x\,T^{00}(t,\mathbf{x})\,x^i \\
    \label{mij}
    M^{ij}(t)&=&\frac{m_I}{m_R}\,\frac{1}{c^2}\int\mathrm{d}^3x\,T^{00}(t,\mathbf{x})\,x^ix^j \\
    \label{mijk}
    M^{ijk}(t)&=&\frac{m_I}{m_R}\,\frac{1}{c^2}\int\mathrm{d}^3x\,T^{00}(t,\mathbf{x})\,x^ix^jx^k~,
\end{eqnarray}
where the QEP corrections appear as modifications to the speed of light $c$. It turns out there is a relation between $S^{ij}$ and $M^{ij}$ as \cite{Maggiore:2007ulw}
\begin{eqnarray}
    S^{ij}\!(t)=\frac{1}{2}\Ddot{M}^{ij}\!(t)~,
\end{eqnarray}
which we use to rewrite Eq. (\ref{httsmom}), considering only the quadrupole contribution, and neglecting higher multipoles, as
\begin{eqnarray}
\label{httmmom}
    h_{ij}^{TT}(t,\mathbf{x})=-\frac{m_Gm_I^{2}}{m_R^3}\,\frac{2G}{c^4}\,\frac{1}{r}\,\Lambda_{ij,kl}(\hat{\mathbf{n}})\,\,\Ddot{M}^{kl}(t_{ret})~.
\end{eqnarray}
To continue the discussion in terms of the quadrupole moment, it is useful to decompose $M^{ij}$ into irreducible representations as \cite{Maggiore:2007ulw}
\begin{eqnarray}
    M^{ij}=\left(M^{ij}-\frac{1}{3}\,\delta^{ij}M_{kk}\right)+\frac{1}{3}\,\delta^{ij}M_{kk}~,
\end{eqnarray}
where the last term can be neglected since $\Lambda_{ij,kl}\,\delta^{kl}=0$. We can then define the QEP modified quadrupole moment as
\begin{eqnarray}
\label{quadmoment}
    Q^{ij}\,(t)\equiv M^{ij}\,(t)-\frac{1}{3}\,\delta^{ij}M_{kk}(t)=\frac{m_R}{m_I}\int\mathrm{d}^3x\,\,\rho(t,\mathbf{x})\left(x^ix^j-\frac{1}{3}\,r^2\delta^{ij}\right)~,
\end{eqnarray}
where we used $T^{00}=\rho\frac{m_R^2}{m_I^2}c^2$, since the QEP modified four velocity scalar product reads as $u^\mu u_\mu=\frac{m_R}{m_I}$ \cite{Das:2023cfu}. Eq. (\ref{httmmom}) can then be rewritten as
\begin{eqnarray}
    h_{ij}^{TT}(t,\mathbf{x})&=&-\frac{m_Gm_I^{2}}{m_R^3}\,\frac{2G}{c^4}\,\frac{1}{r}\,\Lambda_{ij,kl}(\hat{\mathbf{n}})\,\,\Ddot{Q}^{kl}(t_{ret}) \nonumber \\
    &=&-\frac{m_Gm_I^{2}}{m_R^3}\,\frac{2G}{c^4}\,\frac{1}{r}\,\Ddot{Q}_{ij}^{TT}(t_{ret})~,
\end{eqnarray}
where $\Ddot{Q}_{ij}^{TT}(t_{ret})=\Lambda_{ij,kl}(\hat{\mathbf{n}})\,\,\Ddot{Q}^{kl}(t_{ret})$. The radiated power per solid angle, obtained from Eq. (\ref{genenmod}), then reads as
\begin{eqnarray}
\label{qeppowom}
    \frac{\mathrm{d}P}{\mathrm{d}\Omega}=\frac{m_R^{3/2}}{m_Gm_I^{1/2}}\,\frac{c^3r^2}{32\pi G}\,\left\langle\Dot{h}_{ij}^{TT}\Dot{h}_{ij}^{TT}\right\rangle=\frac{m_Gm_I^{7/2}}{m_R^{9/2}}\,\frac{G}{8\pi c^5}\,\Lambda_{ij,kl}(\hat{\mathbf{n}})\,\left\langle\dddot{Q}^{ij}\dddot{Q}^{kl}\right\rangle~,
\end{eqnarray}
which can be used to obtain the total radiated power of the source, by integrating the above over the solid angle, as
\begin{eqnarray}
    P=\frac{m_Gm_I^{7/2}}{m_R^{9/2}}\,\frac{G}{5 c^5}\,\left\langle\dddot{Q}_{ij}\dddot{Q}_{ij}\right\rangle~,
\end{eqnarray}
and, in a similar way, the energy spectrum of the source as
\begin{eqnarray}
    \frac{\mathrm{d}E}{\mathrm{d}\omega}=\frac{m_Gm_I^{7/2}}{m_R^{9/2}}\,\frac{G}{5\pi c^5}\,\omega^6\,\tilde{Q}_{ij}(\omega)\,\tilde{Q}_{ij}^*(\omega)~,
\end{eqnarray}
where $\tilde{Q}_{ij}(\omega)$ is the Fourier transform of the quadrupole moment, defined in Eq. (\ref{quadmoment}).
It is convenient to discuss the GW waveforms in terms of moments $M^{ij}$ as given in Eq. (\ref{httmmom}), and to read out the two polarization amplitudes for a wave propagating in the $\hat{z}$ direction from Eq. (\ref{httmmom})  as
\begin{eqnarray}
    h_+(t,\mathbf{x})&=&-\frac{m_Gm_I^{2}}{m_R^3}\,\frac{G}{c^4}\,\frac{1}{r}\left(\Ddot{M}_{11}(t_{ret})-\Ddot{M}_{22}(t_{ret})\right) \\
    h_\times(t,\mathbf{x})&=&-\frac{m_Gm_I^{2}}{m_R^3}\,\frac{2G}{c^4}\,\frac{1}{r}\,\Ddot{M}_{12}(t_{ret})~.
\end{eqnarray}
In an arbitrary direction $\hat{\mathbf{n}}=(\cos{\phi}\sin{\theta},\sin{\phi}\sin{\theta},\cos{\theta})$ in spherical coordinates, the above amplitudes read as
\begin{eqnarray}
\label{hplar}
     h_+(t;\theta,\phi)=&-&\frac{m_Gm_I^{2}}{m_R^3}\,\frac{G}{c^4}\,\frac{1}{r}\left(\Ddot{M}_{11}(t_{ret})\,(\cos^2{\!\phi}-\sin^2{\!\phi}\cos^2{\!\theta})+\Ddot{M}_{22}(t_{ret})\,(\sin^2{\!\phi}-\cos^2{\!\phi}\cos^2{\!\theta})\right. \nonumber \\
     &-&\Ddot{M}_{33}(t_{ret})\,\sin^2{\!\theta}-\Ddot{M}_{12}(t_{ret})\,\sin{2\phi}\,(1+\cos^2{\!\theta})+\Ddot{M}_{13}(t_{ret})\,\sin{\phi}\,\sin{2\theta} \nonumber \\
     &+&\left.\Ddot{M}_{23}(t_{ret})\,\cos{\phi}\,\sin{2\theta}\right) \\
     \label{hxar}
      h_\times(t;\theta,\phi)=&-&\frac{m_Gm_I^{2}}{m_R^3}\,\frac{G}{c^4}\,\frac{1}{r}\left((\Ddot{M}_{11}(t_{ret})-\Ddot{M}_{22}(t_{ret}))\,\sin{2\phi}\,\cos{\theta}+2\Ddot{M}_{12}(t_{ret})\,\cos{2\phi}\,\cos{\theta}\right. \nonumber \\
      &-&\left.2\Ddot{M}_{13}(t_{ret})\,\cos{\phi}\,\sin{\theta}+2\Ddot{M}_{23}(t_{ret})\,\sin{\phi}\,\sin{\theta}\right)~.
\end{eqnarray}
In the above considerations we have shown how the QEP formalism modifies the linearized Einstein equations. Note that all QEP modifications appear as dimensionless prefactors in terms of the three masses $m_\alpha$. At this point, we did not consider a specific nature of the source of gravitational radiation. The above expressions are general and hold for any source of gravitational radiation, up to the quadrupole contribution.

%To study quadrupole radiation from a mass in circular orbit one needs to consider the modified Newton's law of gravity for the two gravitating objects. The force on the first object is
%\begin{eqnarray}
%    F_{21}=\frac{Gm_{1G}m_{2R}}{d^2}~,
%\end{eqnarray}
%and force on the second object
%\begin{eqnarray}
%    F_{12}=-\frac{Gm_{2G}m_{1R}}{d^2}~,
%\end{eqnarray}
%where $d$ is the distance between the objects, $m_{1,2G}$ are the gravitational masses of the objects and $m_{1,2R}$ the rest masses of the objects, since the rest masses have the physical meaning of active gravitational masses. Note that masses $m_{1\alpha}$ and $m_{2\alpha}$ represent masses of the gravitating objects within the source, while masses $m_\alpha$ from previous equations represent a test particle in the background, generated by the source. It turns out that the above definitions of the gravitational forces on the two objects violate third Newton's law
%\begin{eqnarray}
%    F_{12}=-\frac{m_{2G}/m_{2R}}{m_{1G}/m_{1R}}\,F_{21}
%\end{eqnarray}
%in the case when QEP (or at least LPI) is violated. The modified reduced mass of such a system reads as
%\begin{eqnarray}
%\mu(m_{1\alpha},m_{2\alpha'})\equiv\mu_{m}=\frac{m_{1I}\,m_{2I}\,m_{2R}\,m_{1G}}{m_{2I}\,m_{2R}\,m_{1G}+m_{1I}\,m_{1R}\,m_{2G}}~.
%\end{eqnarray}

\section{Inspiral of compact binaries}
\label{sec:binaries}

There are several possible sources of GW radiation, such as rigid body rotation, radial infall into black holes, accelerating masses and inspiraling compact binaries \cite{Maggiore:2007ulw}. It turns out that out of the above, inspiraling compact binaries are currently the only sources which can produce GW radiation with observable magnitudes. Therefore, we focus on such astrophysical scenarios where compact objects, such as neutron stars and black holes merge, and apply the QEP modified formalism, developed in the previous section. In the center of mass system, the matter density is \cite{Maggiore:2007ulw}
\begin{eqnarray}
\rho(t,\mathbf{x})=\mu\,\delta^{(3)}(\mathbf{x}-\mathbf{x}_0(t))~,
\end{eqnarray}
where $\mu$ is the reduced mass of the binary system and $\mathbf{x}_0(t)$ is the time dependent center of mass coordinate. The matter density is then used to construct the energy-momentum tensor $T^{\mu\nu}$ of the source. The $00$ component reads as
\begin{eqnarray}
    T^{00}(t,\mathbf{x})=\frac{m_R^2}{m_I^2}\,c^2\rho(t,\mathbf{x})~,
\end{eqnarray}
where the additional QEP correction $m_R/m_I$ comes again from the fact that $u^\mu u_{\mu}=m_R/m_I$ and that $T^{\mu\nu}\propto u^\mu u^{\nu}$. By plugging the above in Eq. (\ref{mij}), we obtain
\begin{eqnarray}
\label{mijqep}
    M^{ij}(t)=\frac{m_R}{m_I}\,\mu\,x_0^i(t)\,x_0^j(t)~.
\end{eqnarray}
Note that only Eq. (\ref{mij}) is relevant out of Eqs. (\ref{m}), (\ref{mi}), (\ref{mij}) and (\ref{mijk}), since we only consider the dominant quadrupole contribution, as seen in Eqs. (\ref{hplar}) and (\ref{hxar}). For simplicity we choose a circular orbit as the starting point, which can be described by \cite{Maggiore:2007ulw}
\begin{eqnarray}
x_0(t)&=&R\,\cos{\left(\omega_st+\frac{\pi}{2}\right)} \\
y_0(t)&=&R\,\sin{\left(\omega_st+\frac{\pi}{2}\right)} \\
z_0(t)&=&0~,
\end{eqnarray}
where $R$ is the distance between the centers of mass of the two compact objects and $\omega_s$ is the typical angular frequency of the source (orbital frequency in this case), which turns out as
\begin{eqnarray}
\label{sangfreq}
    \omega_s=\sqrt{\frac{m_G}{m_I}}\,\sqrt{\frac{GM_t}{R^3}}~,
\end{eqnarray}
with $M_t$ being the total mass of the binary, which is not considered as a test mass to study the QEP. Note that the QEP corrections in the above correspond to the masses of the test particle, i.e., the detector mirrors and not to any of the masses of the source. This can be interpreted as an effective modification of the gravitational constant $(m_G/m_I)G$, where a massive test particle with $m_\alpha$ interacts with other masses with such a modified gravitational constant in its reference frame. In other words, each massive particle ``experiences'' the Universe in its own way, i.e., with different constants in its reference frame, which depend on its mass. The non-vanishing components of Eq. (\ref{mijqep}) turn out as
\begin{eqnarray}
    M^{11}(t)&=&\frac{m_R}{m_I}\,\mu\, R^2\,\frac{1-\cos{(2\omega_st)}}{2} \\
    M^{22}(t)&=&\frac{m_R}{m_I}\,\mu\, R^2\,\frac{1+\cos{(2\omega_st)}}{2} \\
    M^{12}(t)&=&-\frac{1}{2}\,\frac{m_R}{m_I}\,\mu\,R^2\sin{(2\omega_st)}~,
\end{eqnarray}
and their second derivatives as
\begin{eqnarray}
    \Ddot{M}^{11}(t)&=&-\Ddot{M}^{22}(t)=2\,\frac{m_R}{m_I}\,\mu\, R^2\,\omega_s^2\cos{(2\omega_st)} \\
    \Ddot{M}^{12}(t)&=&2\,\frac{m_R}{m_I}\,\mu\, R^2\,\omega_s^2\sin{(2\omega_st)}~.
\end{eqnarray}
By plugging the above in Eqs. (\ref{hplar}) and (\ref{hxar}), we obtain a QEP modified GW waveform, created by an inspiraling compact binary as
\begin{eqnarray}
\label{hmod+}
    h_+(t;\theta,\phi)&=&-\frac{m_Gm_I}{m_R^2}\,\frac{4G\mu\,\omega_s^2R^2}{c^4}\,\frac{1}{r}\left(\frac{1+\cos^2{\theta}}{2}\right)\cos{(2\omega_st_{ret}+2\phi)}~, \\
    \label{hmodx}
    h_\times(t;\theta,\phi)&=&-\frac{m_Gm_I}{m_R^2}\,\frac{4G\mu\,\omega_s^2R^2}{c^4}\,\frac{1}{r}\,\cos{\theta}\,\sin{(2\omega_st_{ret}+2\phi)}~.
\end{eqnarray}
 The QEP modified radiated power per solid angle, defined in Eq. (\ref{qeppowom}), in the case of inspiraling compact binaries, and expanded to the quadrupole contribution, now reads as
\begin{eqnarray}
\label{radpowsolang}
    \frac{\mathrm{d}P}{\mathrm{d}\Omega}=\frac{m_R^{3/2}}{m_Gm_I^{1/2}}\,\frac{c^3r^2}{16\pi G}\,\left\langle\Dot{h}_+^2+\Dot{h}_\times^2\right\rangle=\frac{m_Gm_I^{3/2}}{m_R^{5/2}}\,\frac{2G\mu^2R^4\omega_s^6}{\pi c^5}\left(\left(\frac{1+\cos^2{\theta}}{2}\right)^{\!2}+\cos^2{\theta}\right)~,
\end{eqnarray}
and by integrating the above over the solid angle, the total radiated power as
\begin{eqnarray}
\label{radpow}
    P=\frac{m_Gm_I^{3/2}}{m_R^{5/2}}\,\frac{32G\mu^2R^4\omega_s^6}{c^5}~.
\end{eqnarray}

It is useful to introduce the chirp mass of the source as $M_c=\mu^{3/5}M_t^{2/5}$ \cite{Maggiore:2007ulw}, as the mass parameter of the binary system. The chirp mass and Eq. (\ref{sangfreq}) can then be used to rewrite Eqs. (\ref{hmod+}) and (\ref{hmodx}) as
\begin{eqnarray}
\label{halmost+}
    h_+(t;\theta,\phi)&=&-\frac{m_G^{5/3}m_I^{1/3}}{m_R^2}\,4\left(\frac{GM_c}{c^2}\right)^{5/3}\left(\frac{\omega_s}{c}\right)^{2/3}\frac{1}{r}\left(\frac{1+\cos^2{\theta}}{2}\right)\cos{(2\omega_st_{ret}+2\phi)}~, \\
    \label{halmostx}
    h_\times(t;\theta,\phi)&=&-\frac{m_G^{5/3}m_I^{1/3}}{m_R^2}\,4\left(\frac{GM_c}{c^2}\right)^{5/3}\left(\frac{\omega_s}{c}\right)^{2/3}\frac{1}{r}\cos{\theta}\sin{(2\omega_st_{ret}+2\phi)}~,
\end{eqnarray}
and Eqs. (\ref{radpowsolang}) and (\ref{radpow}) as
\begin{eqnarray}
    \frac{\mathrm{d}P}{\mathrm{d}\Omega}&=&\frac{m_G^{7/3}m_I^{1/6}}{m_R^{5/2}}\,\frac{2}{\pi}\,\frac{c^5}{G}\left(\frac{GM_c\omega_{_{GW}}}{2 c^3}\right)^{\!10/3}\left(\left(\frac{1+\cos^2{\theta}}{2}\right)^{\!2}+\cos^2{\theta}\right)~, \\
    P&=&\frac{m_G^{7/3}m_I^{1/6}}{m_R^{5/2}}\,\frac{32}{5}\,\frac{c^5}{G}\left(\frac{GM_c\omega_{_{GW}}}{2c^3}\right)^{\!10/3}~,
\end{eqnarray}
where $\omega_{_{GW}}=2\omega_s$, which implies that the angular frequency of the GW is twice that of the source \cite{Maggiore:2007ulw}. In the Newtonian limit, the energy of the orbit  in the binary system can be written as
\begin{eqnarray}
    E_{b}=-\frac{m_G^{2/3}}{m_I^{2/3}}\left(\frac{G^2M_c^5\omega_{_{GW}}^2}{32}\right)^{\!1/3}~,
\end{eqnarray}
where the QEP correction appears as an effective correction to the gravitational constant, as observed by the massive detector mirror. As the gravitational radiation decreases the orbital energy of the binary system, the time-dependent gravitational wave frequency can be obtained by \cite{Maggiore:2007ulw}
\begin{eqnarray}
    \frac{\mathrm{d}E_b}{\mathrm{d}t}=-P~,
\end{eqnarray}
from which we obtain the differential equation
\begin{eqnarray}
    \frac{\mathrm{d}f_{_{GW}}}{\mathrm{d}t}=\frac{m_G^{5/3}m_I^{5/6}}{m_R^{5/2}}\,\frac{96}{5}\,\pi^{8/3}\left(\frac{GM_c}{c^3}\right)^{\!5/3}f_{_{GW}}^{11/3}~,
\end{eqnarray}
where $\omega_{_{GW}}=2\pi f_{_{GW}}$. We then integrate the above from some initial time $t$ and frequency $f_{_{GW}}(t)$ to the time at coalescence $t_c$ and frequency $f_{_{GW}}(t_c)\longrightarrow\infty$ to obtain
\begin{eqnarray}
\label{modinspfreq}
    f_{_{GW}}(\tau)=\frac{m_R^{15/16}}{m_G^{5/8}m_I^{5/16}}\,\frac{1}{\pi}\left(\frac{5}{256}\,\frac{1}{\tau}\right)^{\!3/8}\left(\frac{c^3}{GM_c}\right)^{\!5/8}~,
\end{eqnarray}
where $\tau=t_{c}-t$ is the time until coalescence of the binary at time $t$, which, after a rearrangement of the above equation, reads as
\begin{eqnarray}
    \tau=\frac{m_R^{5/2}}{m_G^{5/3}m_I^{5/6}}\,\frac{5}{256}\,\frac{1}{\pi^{8/3}}\left(\frac{c^3}{GM_c}\right)^{\!5/3}\frac{1}{f_{_{GW}}^{8/3}}~.
\end{eqnarray}
In the above, we can interpret $f_{_{GW}}$ as the gravitational wave frequency at which the detector initially picked up the GW signal. The phase of the waveform $\Phi$ is then obtained by simply integrating $\omega_{_{GW}}(t)$ over time
\begin{eqnarray}
   \label{modphase} \int_{\Phi_c}^{\Phi(t)}\!\!\mathrm{d}\Phi=\int_{t_c}^t\mathrm{d}t'\,\omega_{_{GW}}(t')\,\,\,\,\,\,\,\,\implies\,\,\,\,\,\,\,\,\Phi(\tau)=-2\,\frac{m_R^{15/16}}{m_G^{5/8}m_I^{5/16}}\left(\frac{c^3}{5GM_c}\right)^{\!5/8}\tau^{5/8}+\Phi_c~,
\end{eqnarray}
where $\Phi_c$ is the phase at coalescence. 
%Note that the above results are given as measured in the detector frame, using a test particle with masses $m_\alpha$. The actual physical quantities of the source (such as radiated power, orbital frequency, coalescence time and phase), when not directly measured, are independent of the test particle / detector, and therefore do not include modifications.
By plugging Eqs. (\ref{modinspfreq}) and (\ref{modphase}) in Eqs. (\ref{halmost+}) and (\ref{halmostx}), we obtain
\begin{eqnarray}
    \label{hthere+}
    h_+(t)&=&-\frac{m_G^{5/4}m_I^{1/8}}{m_R^{11/8}}\left(\frac{GM_c}{c^2}\right)^{\!5/4}\left(\frac{5}{c\tau}\right)^{\!1/4}\frac{1}{r}\left(\frac{1+\cos^2{\iota}}{2}\right)\cos{\Phi(\tau)}~, \\
    \label{htherex}
    h_\times(t)&=&-\frac{m_G^{5/4}m_I^{1/8}}{m_R^{11/8}}\left(\frac{GM_c}{c^2}\right)^{\!5/4}\left(\frac{5}{c\tau}\right)^{\!1/4}\frac{1}{r}\cos{\iota}\,\sin{\Phi(\tau)}~,
\end{eqnarray}
which are now expressed in terms of $\tau$. In the above, the angle $\theta$ is identified as the inclination of the orbital plane $\iota$ of the binary system and the coordinate system of the binary is chosen such that the Earth is located in its $x-z$ plane, where $\phi=0$.

The radiated energy can be measured with GW detectors. Therefore, we want to see how it gets modified in order to test the EEP. To achieve this, we take the general expression for the energy per solid angle from the first line of Eq. (\ref{genenmod}) and rewrite it as the energy spectrum as
\begin{eqnarray}
\label{modenspec}
    \frac{\mathrm{d}E}{\mathrm{d}f}=\frac{m_R^{3/2}}{m_Gm_I^{1/2}}\,\frac{\pi c^3 f^2r^2}{2G}\int\mathrm{d}\Omega
\left(\left|\tilde{h}_{+}(f)\right|^2+\left|\tilde{h}_{\times}(f)\right|^2\right)~,
\end{eqnarray}
where \cite{Maggiore:2007ulw}
\begin{eqnarray}
\label{hfourier}
    \tilde{h}_{+,\times}(f)=\int \mathrm{d}t\,h_{+,\times}(t)\,e^{i2\pi ft}\simeq\frac{1}{2}\,e^{i\psi_{+,\times}(f)}\,A_{+,\times}(t_*)\left(\frac{2}{\ddot{\Phi}(t_*)}\right)^{\!1/2}~
\end{eqnarray}
are the Fourier transforms of $h_{+,\times}$, evaluated at the stationary point $t_*$, determined by $2\pi f=\dot{\Phi}(t_*)$. In the above
\begin{eqnarray}
    \psi_{+}(f)&=&2\pi f \left(t_c-\sqrt{\frac{m_I}{m_R}}\,\frac{r}{c}\right)-\Phi_c-\frac{\pi}{4}+\frac{m_R^{5/2}}{m_G^{5/3}m_I^{5/6}}\,\frac{3}{4}\left(\frac{c^3}{GM_c}\,\frac{1}{8\pi f}\right)^{\!5/3}~, \\
     \psi_{\times}(f)&=&\psi_{+}(f)+\frac{\pi}{2}~,
\end{eqnarray}
and
\begin{eqnarray}
    A_+(t)&=&-\frac{m_G^{5/4}m_I^{1/8}}{m_R^{11/8}}\left(\frac{GM_c}{c^2}\right)^{\!5/4}\left(\frac{5}{c\tau}\right)^{\!1/4}\frac{1}{r}\left(\frac{1+\cos^2{\iota}}{2}\right)~, \\
    A_\times(t)&=& -\frac{m_G^{5/4}m_I^{1/8}}{m_R^{11/8}}\left(\frac{GM_c}{c^2}\right)^{\!5/4}\left(\frac{5}{c\tau}\right)^{\!1/4}\frac{1}{r}\,\cos{\iota}~.
\end{eqnarray}
In order to obtain the modified energy spectrum from Eq. (\ref{modenspec}), we first need to obtain the modified Fourier transforms from Eq. (\ref{hfourier}). By plugging Eqs. (\ref{hthere+}) and (\ref{htherex}) in Eq. (\ref{hfourier}) we obtain
\begin{eqnarray}
\label{hfour+}
    \tilde{h}_{+}(f)&=&-\frac{m_G^{5/6}}{m_R^{3/4}m_I^{1/12}}\,\frac{1}{\pi^{2/3}}\left(\frac{5}{24}\right)^{\!1/2}e^{i\psi_{+}(f)}\,\frac{c}{r}\left(\frac{GM_c}{c^3}\right)^{\!5/6}\frac{1}{f^{7/6}}\left(\frac{1+\cos^2{\iota}}{2}\right)~, \\
    \label{hfourx}
    \tilde{h}_{\times}(f)&=&-\frac{m_G^{5/6}}{m_R^{3/4}m_I^{1/12}}\,\frac{1}{\pi^{2/3}}\left(\frac{5}{24}\right)^{\!1/2}e^{i\psi_{\times}(f)}\,\frac{c}{r}\left(\frac{GM_c}{c^3}\right)^{\!5/6}\frac{1}{f^{7/6}}\,\cos{\iota}~.
\end{eqnarray}
The modified energy spectrum from Eq. (\ref{modenspec}) then reads as
%By plugging the above Fourier transforms in Eq. (\ref{modenspec}), one obtains
\begin{eqnarray}
    \frac{\mathrm{d}E}{\mathrm{d}f}=\frac{m_G^{2/3}}{m_I^{2/3}}\,\frac{\pi^{2/3}}{3G}\,(GM_c)^{5/3}\frac{1}{f^{1/3}}~.
\end{eqnarray}
%In the above, notice that the rest mass dependence cancels. This makes sense since the energy spectrum of the source is not dependent on the speed of information transfer (causality) $c'=c\sqrt{\frac{m_R}{m_I}}$ between the source and the detector. 
By integrating the above energy spectrum over the frequency range $(f_{min},f_{max})$, where $f_{min}$ is the lowest frequency in the detection history of the source, and $f_{max}$ the frequency at coalescence, where $f_{min}\ll f_{max}$, such that $f_{min}$ can be dropped in the following, we obtain the total radiated energy as
%\textcolor{red}{\bf Are we saying that the Detector is affecting the emitted power. Is it? In principle, 2 merging stars should produce GW, why the rate of GW energy emission is affected by the detector masses $m_G/m_R$? It is like Brans Dicke theory in which the far mass distribution affects locally the gravitational physics through, for example, a varying Gravitational coupling constant. is it?}
\begin{eqnarray}
\label{raden}
    \Delta E_{rad}=\frac{m_G^{2/3}}{m_I^{2/3}}\,\frac{\pi^{2/3}}{2G}\,(GM_c)^{5/3}f_{max}^{2/3}~.
\end{eqnarray}
To obtain $f_{max}$, we need to work out the radius at which the objects coalesce. This happens when strong gravitational effects take over, and circular motion is not allowed. This is quantified by the so-called Innermost Stable Circular Orbit (ISCO), which reads as
\begin{eqnarray}
    R_{ISCO}=\frac{m_G}{m_R}\,\frac{6GM_t}{c^2}~
\end{eqnarray}
in the case of the Schwarzschild space-time. 
%Note that the above introduces new corrections since it represents a quantity which depends on the speed of information (causality) between the source and the detector. 
By plugging $R_{ISCO}$ in Eq. (\ref{sangfreq}), we obtain
\begin{eqnarray}
    f_{max}\equiv (f_s)_{ISCO}=\frac{m_R^{3/2}}{m_Gm_I^{1/2}}\,\frac{1}{12\sqrt{6}\,\pi}\,\frac{c^3}{GM_t}~,
\end{eqnarray}
and by plugging the above in Eq. (\ref{raden}), we obtain the total radiated energy as
\begin{eqnarray}
    \Delta E_{rad}=\frac{m_R}{m_I}\,\frac{1}{12}\,\mu c^2~,
\end{eqnarray}
which depends only on the reduced mass of the source and ratio $m_R/m_I$ of the detector. A more precise derivation, considering curved space-time at all steps, results in 
\begin{eqnarray}
\label{predraden}
     \Delta E_{rad}=\frac{m_R}{m_I}\left(1-\sqrt{\frac{8}{9}}\right)\mu c^2~,
\end{eqnarray}
which corresponds to the binding energy of an ISCO. Note that the QEP modifications suggest a modified speed of light and gravitational constant as seen by the massive detector and \emph{do not} directly affect the physical process of generating GWs. This covers the formulation of the QEP modified GW theory for compact inspiraling binaries. Such modifications are an essential ingredient in testing the EEP with actual experiments, as shown in the following section.

\section{Bounds on the violation from the observed gravitational wave events}
\label{sec:bounds}

In this section, we use the QEP modified GW theory for the inspiral of compact binaries, obtained in the previous section, to test the three aspects of the EEP, namely WEP, LLI and LPI, using the LIGO/Virgo GW detectors. As in the standard theory, the QEP modified GW theory is used to construct observables, which are then compared to observations of actual GW events. It is convenient to define a scalar GW signal, which is the output of the detector and can be written as \cite{Maggiore:2007ulw}
\begin{eqnarray}
    h(t)=D^{ij}h_{ij}(t)~,
\end{eqnarray}
where $D^{ij}$ is the so-called detector tensor and depends on the geometry of the detector. The above scalar GW signal can also be written in terms of pattern functions as \cite{Maggiore:2007ulw}
\begin{eqnarray}
\label{htotdet}
    h(t)=h_{+}\!(t)\,F_{+}\!(\theta,\phi)+h_{\times}\!(t)\,F_{\times}\!(\theta,\phi)~,
\end{eqnarray}
where $F_{+,\times}\!(\hat{\mathbf{n}})=D^{ij}\,e_{ij}^{+,\times}\!(\hat{\mathbf{n}})$ are the form factors, with polarization tensors $e_{ij}^{+,\times}\!(\hat{\mathbf{n}})$. Note that here, $\hat{\mathbf{n}}$ or $\theta$ and $\phi$ refer to the direction of the source in the sky, relative to the detector, and not to the relative orientation of the orbit of the binary as seen in Eqs. (\ref{halmost+}) and (\ref{halmostx}). In the specific case of the LIGO/Virgo detectors, the detector geometry is that of an interferometer, for which the form factors read as \cite{Maggiore:2007ulw}
\begin{eqnarray}
\label{form+}
    F_{+}\!(\theta,\phi)&=&\frac{1+\cos^2{\theta}}{2}\,\cos{2\phi}~, \\
    \label{formx}
     F_{\times}\!(\theta,\phi)&=&\cos{\theta}\,\sin{2\phi}~.
\end{eqnarray}

Here, we take two observables into consideration, namely the GW signal and the time dependence of the orbital frequency of the inspiraling binary. First, considering the GW signal in Eq. (\ref{htotdet}), using the QEP modified waveform from Eqs. (\ref{hthere+}) and (\ref{htherex}) and the expectation values $\langle \hat{H}_{int,\alpha}\rangle\equiv E_\alpha$, we obtain a bound on the general EEP violation as (see Appendix \ref{qeptests})
\begin{eqnarray}
\label{qepgen1}
    \mathrm{LB}_A<\frac{1}{mc^2}\left(\frac{5}{4}E_G+\frac{1}{8}E_I-\frac{11}{8}E_R\right)<\mathrm{UB}_{A}~,
\end{eqnarray}
%\begin{eqnarray}
%\label{qepgen1}
%    \mathrm{LB}_S<\frac{1}{mc^2}\left(\frac{5}{3}E_G-2E_R+\frac{1}{3}E_I\right)<\mathrm{UB}_{S}~,
%\end{eqnarray}
where $\mathrm{LB}_{A}$ and $\mathrm{UB}_{A}$ are the relative lower and upper bounds, respectively, and are obtained from the GW amplitude signal measurements. Next, considering the QEP modified time dependence of the orbital frequency in Eq. (\ref{modinspfreq}), we obtain another bound on the general EEP violation as (see Appendix \ref{qeptests})
\begin{eqnarray}
\label{qepgen2}
    \mathrm{LB}_f<\frac{1}{mc^2}\left(\frac{15}{16}E_R-\frac{5}{8}E_G-\frac{5}{16}E_I\right)<\mathrm{UB}_f~,
\end{eqnarray}
where $\mathrm{LB}_{f}$ and $\mathrm{UB}_{f}$ are the relative lower and upper bounds, respectively, and are obtained from the time dependence of the orbital frequency measurements. Bounds obtained from specific GW events are shown in Table \ref{tab:bounds}. We have obtained these bounds from the current three most resolved GW events (see Appendix \ref{boundpendix}), detected by the LIGO/Virgo detectors, which include GW170817 \cite{Abbot:2020a}, GW190521 \cite{Abbot:2020b} and GW190814 \cite{Abbot:2020}.
\begin{table}[H]
    \centering
    \begin{tabular}{|c|c|c|c|}
    \hline
        {}     &  GW170817 & GW190521 & GW190814 \\
        \hline
       $LB_A$  & $-0.350$ & $-0.53$ & $-0.187$  \\
       \hline
       $UB_A$  & $0.200$ & $0.51$ & $0.171$ \\
       \hline
       $LB_f$  & $-0.00016$ & $-0.096$ & $-0.0062$  \\
       \hline
       $UB_f$  & $0.00042$ & $0.115$ & $0.0062$ \\
       \hline
    \end{tabular}
    \caption{Bounds obtained from GW events}
    \label{tab:bounds}
\end{table}
Note that Eqs. (\ref{qepgen1}) and (\ref{qepgen2}) include all three expectation values of internal energy, i.e., $E_I$, $E_G$ and $E_R$, while the individual tests of WEP, LLI and LPI require only specific pairs of these expectation values. The individual tests for WEP, LLI and LPI, can then be constructed by combining Eqs. (\ref{qepgen1}) and (\ref{qepgen2}) in three different ways, such that one of the expectation values cancels out. The WEP test condition turns out as
\begin{eqnarray}
\label{qepv1}
    \frac{22}{5}\,\mathrm{LB}_f+3\,\mathrm{LB}_A%=\left\{\begin{array}{c}
        % -1.05  \\
       %  -2.01  \\
      %   -0.59
    %\end{array}\right\}
    <\frac{1}{mc^2}\,(E_G-E_I)<%\left\{\begin{array}{c}
         %0.60  \\
        % 2.04  \\
        % 0.54
    %\end{array}\right\}=
    \frac{22}{5}\,\mathrm{UB}_f+3\,\mathrm{UB}_A~,
\end{eqnarray}
the LLI test condition as
\begin{eqnarray}
\label{qepv2}
    4\,\mathrm{LB}_f+2\,\mathrm{LB}_A%=\left\{\begin{array}{c}
      %   -0.70  \\
       %  -1.44  \\
      %  -0.40
    %\end{array}\right\}
    <\frac{1}{mc^2}\,(E_R-E_I)<%\left\{\begin{array}{c}
        % 0.40  \\
       %  1.48  \\
       %  0.37
    %\end{array}\right\}=
    4\,\mathrm{UB}_f+2\,\mathrm{UB}_A~,
\end{eqnarray}
and the LPI test condition as
\begin{eqnarray}
\label{qepv3}
    \frac{2}{5}\,\mathrm{LB}_f+\mathrm{LB}_A<%\left\{\begin{array}{c}
      %   -0.35  \\
       %  -0.57  \\
        % -0.19
    %\end{array}\right\}=
    \frac{1}{mc^2}\,(E_G-E_R)<%\left\{\begin{array}{c}
      %   0.20  \\
       %  0.56  \\
       %  0.17
    %\end{array}\right\}=
    \frac{2}{5}\,\mathrm{UB}_f+\mathrm{UB}_A~.
\end{eqnarray}
Note that the lower and upper bounds on the three aspects of the EEP are now linear combinations of the relative precisions of both the amplitude and frequency measurements. The specific bounds on the three aspects of the EEP are shown in Table \ref{tab:qepbounds}.

\begin{table}[H]
    \centering
    \begin{tabular}{|c||c|c|c||c|c|c||c|c|c|}
    \hline
    Source & \multicolumn{3}{|c||}{WEP test} & \multicolumn{3}{|c||}{LLI test} & \multicolumn{3}{|c|}{LPI test} \\
    \hline
        GW170817  & $-1.05$ & {} & $0.60$ & $-0.70$ & {} & $0.40$ & $-0.35$ & {} & $0.20$ \\
        %\hline
        GW190521  & $-2.01$ & $\displaystyle{<\frac{1}{mc^2}\,(E_G-E_I)<}$ & $2.04$ & $-1.44$ & $\displaystyle{<\frac{1}{mc^2}\,(E_R-E_I)<}$ & $1.48$ & $-0.57$ & $\displaystyle{<\frac{1}{mc^2}\,(E_G-E_R)<}$ & $0.56$  \\
       %\hline
        GW190814  & $-0.59$ & {} & $0.54$ & $-0.40$ & {} & $0.37$ & $-0.19$ & {} & $0.17$ \\
       \hline
    \end{tabular}
    \caption{EEP bounds}
    \label{tab:qepbounds}
\end{table}

%\begin{table}[H]
%    \centering
%    \begin{tabular}{|c|c|c|c|}
%    \hline
%    \multicolumn{4}{|c|}{WEP test} \\ %& & & \\
%    \hline
%        GW170817  & $-1.05$ & {} & $0.60$ \\
%        %\hline
%        GW190521  & $-2.01$ & $\displaystyle{<\frac{1}{mc^2}\,(E_G-E_I)<}$ & $2.04$  \\
%       %\hline
%        GW190814  & $-0.59$ & {} & $0.54$ \\
%       \hline
%       \multicolumn{4}{|c|}{LLI test} \\ %& & & \\
%    \hline
%        GW170817  & $-0.70$ & {} & $0.40$ \\
%        %\hline
%        GW190521  & $-1.44$ & $\displaystyle{<\frac{1}{mc^2}\,(E_R-E_I)<}$ & $1.48$  \\
%       %\hline
%        GW190814  & $-0.40$ & {} & $0.37$ \\
%       \hline
%       \multicolumn{4}{|c|}{LPI test} \\ %& & & \\
%    \hline
%        GW170817  & $-0.35$ & {} & $0.20$ \\
%        %\hline
%        GW190521  & $-0.57$ & $\displaystyle{<\frac{1}{mc^2}\,(E_G-E_R)<}$ & $0.56$  \\
%       %\hline
%        GW190814  & $-0.19$ & {} & $0.17$ \\
%       \hline
%    \end{tabular}
%    \caption{QEP tests}
%    \label{tab:bounds}
%\end{table}

There are also other observables which can provide bounds on the EEP but are, in general, less precise than the above consideration. We can use the radiated energy and the signal-to-noise measurement for the events. For example, given a measurement of radiated energy for the GW170817 event, $\Delta E_{rad}^{GW170817}=(6.3\pm1.8)\times10^{47}\,\mathrm{J}$ \cite{refId0}, one can obtain the bound for the LLI from Eq. (\ref{predraden}) as (see Appendix \ref{qeptests})
\begin{eqnarray}
    \left|\frac{1}{mc^2}\,(E_R-E_I)\right|<0.286~.
\end{eqnarray}
Another potential avenue for obtaining the EEP bounds is the signal-to-noise ratio, which turns out not to be as useful as the above approach. However, we present the results below for completeness. The signal-to-noise ratio is defined as \cite{Maggiore:2007ulw}
\begin{eqnarray}
\label{snr}
    \frac{S}{N}=2\sqrt{\int_0^\infty \!\mathrm{d}f\,\frac{|\tilde{h}(f)|^2}{S_n(f)}}~,
\end{eqnarray}
where
\begin{eqnarray}
    \tilde{h}(f)=F_{+}\!(\hat{\mathbf{n}})\,\tilde{h}_{+}\!(f)+F_{\times}\!(\hat{\mathbf{n}})\,\tilde{h}_{\times}\!(f)
\end{eqnarray}
is the Fourier transform of $h(t)$ and 
\begin{eqnarray}
    S_n(f)=2\int_{-\infty}^\infty\!\mathrm{d}\tau\,\langle\, n(t+\tau)\,n(t)\rangle\, e^{i2\pi f\tau}
\end{eqnarray}
the square of the spectral strain density, with $n(t)$ being the detector noise. The noise in the LIGO/Virgo detectors is primarily comprised of shot noise and radiation pressure fluctuations. The modified shot noise contribution turns out as
\begin{eqnarray}
\label{snshot}
    \left.S_n^{1/2}(f)\right|_{\mathrm{shot}}=\frac{m_R^{1/4}}{m_I^{1/4}}\,\frac{1}{8\mathcal{F}L}\left(\frac{4\pi \hbar\lambda_Lc}{\eta P_{bs}}\right)^{\!1/2}\sqrt{1+\left({f}/{f_p}\right)^2}~,
\end{eqnarray}
where $\mathcal{F}$ is known as the finesse factor of a Fabry-Perot cavity, $L$ the interferometer arm length, $\lambda_L$ the laser wavelength, $\eta$ the efficiency of the photodetector, $P_{bs}$ the laser power on the beamsplitter and $f_p=1/4\pi\tau_s$ the pole frequency, with $\tau_s$ being the storage time of the Fabry-Perot cavity.
The radiation pressure contribution turns out as
\begin{eqnarray}
\label{snradiation}
    \left.S_n^{1/2}(f)\right|_{\mathrm{rad}}=\frac{1}{m_R^{1/4}m_I^{3/4}}\,\frac{16\sqrt{2}\mathcal{F}}{L(2\pi f)^2}\,\sqrt{\frac{\hbar}{2\pi}\,\frac{P_{bs}}{\lambda_Lc}}\,\,\frac{1}{\sqrt{1+\left(f/f_p\right)^2}}~.
\end{eqnarray}
Note that we obtained the QEP modifications in Eqs. (\ref{snshot}) and (\ref{snradiation}) by inserting the effective speed of light correction $c'=c\sqrt{m_R/m_I}$ in the standard equations from Ref. \cite{Maggiore:2007ulw}, instead of $c$ and identifying the explicit mass of the mirror in Eq. (\ref{snradiation}) as the inertial mass $m_I$, while assuming that $\hbar$ obtains no QEP modifications.
The total spectral strain density is then the sum $S_n(f)=\left.S_n(f)\right|_{\mathrm{shot}}+\left.S_n(f)\right|_{\mathrm{rad}}$. The noise is minimized in the standard quantum limit (SQL) as 
\begin{eqnarray}
\label{noisestrain}
    \left.S_n^{1/2}(f)\right|_{\mathrm{SQL}}=\frac{1}{2\pi L f}\,\sqrt{\frac{8\hbar}{m_I}}~,
%\frac{m_R^{1/2}}{m_I}\frac{\sqrt{8\hbar}}{2\pi fL}~.
\end{eqnarray}
which is obtained by finding the optimal frequency
\begin{eqnarray}
f_0=\frac{1}{m_R^{1/2}}\,\frac{8\mathcal{F}}{2\pi}\,\sqrt{\frac{P_{bs}}{\pi\lambda_Lc}}~,    
\end{eqnarray}
where the contributions from the shot noise and radiation pressure noise are equal, i.e.,
\begin{eqnarray}
    1+\left(\frac{f}{f_p}\right)^{\!2}=\frac{m_R^{1/2}}{m_I^{1/2}}\,\frac{f_0^2}{f^2}~.
\end{eqnarray}
By plugging Eqs. (\ref{hfour+}), (\ref{hfourx}), (\ref{form+}), (\ref{formx}) and (\ref{noisestrain}) in Eq. (\ref{snr}), we obtain
\begin{eqnarray}
\label{sigtonoise}
    \frac{S}{N}=\frac{m_G^{5/6}m_I^{5/12}}{m_R^{3/4}}\,\sqrt{\frac{5}{8}\,\frac{\pi^{2/3}L^2}{\hbar}\,\frac{c^2}{r^2}\left(\frac{GM_c}{c^3}\right)^{\!5/3}\!f_{max}^{2/3}\,\left|Q(\theta,\phi;\iota)\right|^2}~,
\end{eqnarray}
where
\begin{eqnarray}
    Q(\theta,\phi;\iota)=\frac{1+\cos^2{\theta}}{2}\,\cos{2\phi}\,\frac{1+\cos^2{\iota}}{2}+i\cos{\theta}\,\sin{2\phi}\,\cos{\iota}~.
\end{eqnarray}
Now, considering the QEP modifications in Eq. (\ref{sigtonoise}) and the precision of the signal-to-noise ratio measurement, we obtain a bound on another general EEP violation as (see Appendix \ref{qeptests})
\begin{eqnarray}
    \,\mathrm{LB}_{SNR}<\frac{1}{mc^2}\,\left(\frac{5}{6}E_G-\frac{3}{4}E_R+\frac{5}{12}E_I\right)<\,\mathrm{UB}_{SNR}~,
\end{eqnarray}
where $\mathrm{LB}_{SNR}$ and $\mathrm{UB}_{SNR}$ are the relative lower and upper bounds, respectively, obtained from the signal-to-noise ratio. We used the above general EEP violation condition alongside the scalar signal amplitude general EEP violation condition from Eq. (\ref{qepgen1}) to construct the bounds on the WEP as
\begin{eqnarray}
    \frac{33\,\mathrm{LB}_{SNR}-18\,\mathrm{LB}_{A}-\frac{33}{2}\,\frac{E_I}{mc^2}}{5}<\frac{1}{mc^2}\,(E_G-E_I)<\frac{33\,\mathrm{UB}_{SNR}-18\,\mathrm{UB}_{A}-\frac{33}{2}\,\frac{E_I}{mc^2}}{5}~,
\end{eqnarray}
on the LLI as
\begin{eqnarray}
    {6\,\mathrm{LB}_{SNR}-4\,\mathrm{LB}_{A}-{3}\,\frac{E_I}{mc^2}}<\frac{1}{mc^2}\,(E_R-E_I)<{6\,\mathrm{UB}_{SNR}-4\,\mathrm{UB}_{A}-{3}\,\frac{E_I}{mc^2}}~,
\end{eqnarray}
and on the LPI as
\begin{eqnarray}
    \frac{20\,\mathrm{LB}_{A}-6\,\mathrm{LB}_{SNR}+3\,\frac{E_R}{mc^2}}{20}<\frac{1}{mc^2}\,(E_G-E_R)<\frac{20\,\mathrm{UB}_{A}-6\,\mathrm{UB}_{SNR}+3\,\frac{E_R}{mc^2}}{20}~.
\end{eqnarray}
Note that there are terms with $E_I$ and $E_R$ in the above expressions for the bounds. They present a significant problem since we would have to know their exact values to obtain useful numerical values for the bounds. They appear because the spectral strain density explicitly depends on the inertial mass of the mirror. Therefore, it is more convenient to test the EEP using the bounds obtained in Eqs. (\ref{qepv1}), (\ref{qepv2}) and (\ref{qepv3}).
%is $S/N=25.0_{-0.2}^{+0.1}$ \cite{Abbot:2020}, for the GW190814 event.

\section{Conclusion}
\label{sec:conc}

In this work, we have provided a general theoretical framework for testing the EEP with GW detectors. We have obtained QEP modifications to the linearized Einstein equations and their solutions, as part of which we had to find the QEP modified Green's function. The QEP modifications, derived in section \ref{sec:mods}, are generally valid for any type of GW source and any type of GW detector. In section \ref{sec:binaries}, we have chosen a specific type of GW source, namely inspiraling compact binaries, since they are the most common and prominent sources of gravitational radiation in the Universe. In the same way, as seen in Ref. \cite{Das:2023cfu}, it turns out that the QEP modifications appear as effective modifications to the speed of light $c'=c\sqrt{m_R/m_I}$ and the gravitational constant $G'=G\,m_G/m_I$. %Note that the main theoretical framework is independent of the GW detector type.
Note that we have considered expectation values of $\hat{M}_\alpha$ to provide phenomenological predictions, which means that the tests refer to the classical counterpart of the QEP, namely the EEP.

In section \ref{sec:bounds}, we have constructed observables which can be observed in an interferometer GW detector, such as LIGO/Virgo, KAGRA and LISA. The constructed observables include the time-dependent inspiral frequency, the waveform, radiated energy and the signal-to-noise ratio, as seen in Eqs. (\ref{modinspfreq}), (\ref{hthere+}), (\ref{htherex}), (\ref{predraden}) and (\ref{sigtonoise}), respectively. Note that it is not convenient to test the EEP using the signal-to-noise ratio since the bounds include the internal energy contributions, which would need to be known to obtain a valid bound.

As an example of a EEP test with a GW detector, we have chosen the three most resolved GW events detected by the LIGO/Virgo detectors, namely GW170817, GW190521 and GW190814. We tested the three features of the EEP, namely WEP, LLI and LPI, as can be seen in Table \ref{tab:qepbounds}. No violations were observed, but we obtained bounds on them. Note that the obtained bounds are not an improvement on the existing bounds on the aspects of the EEP, as can be seen in comparison with Ref. \cite{CMW}. However, the results remain relevant since more precise future GW detectors will have much higher precision and will be able to improve the existing bounds using the formalism proposed here.

%\textcolor{red}{\bf Can we extend/apply the analysis to Microscope experiment, which plans to reach a sensitivity of the order $10^{-15}$}

\section*{Acknowledgement}

This work was supported by the Natural Sciences and Engineering Research Council of Canada.

\section*{Conflict of interest}

The authors declare no conflicting interests.

\section*{Data and code availability}

Data and code sharing are not applicable to this article as no new data were created or analyzed in this study.

 \appendix

 \section{QEP modified energy spectrum}
 \label{app:qepenspec}

 By considering the linearized Einstein equations, the energy-momentum tensor $T_{\mu\nu}$ can be decomposed into the contribution of the source $\Bar{T}_{\mu\nu}$ and the contribution of the gravitational field $t_{\mu\nu}$ as $T_{\mu\nu}=\bar{T}_{\mu\nu}+t_{\mu\nu}$. Away from the source, we can observe that $\Bar{T}_{\mu\nu}=0$, and that $T_{\mu\nu}=t_{\mu\nu}$ \cite{Maggiore:2007ulw}. The energy-momentum tensor is always conserved, which means that away from the source the conservation law reads as $\partial_\mu t^{\mu\nu}=0$, and in the context of the QEP, where $\partial_\mu=\left(\frac{1}{c}\,\partial_0,\sqrt{\frac{m_R}{m_I}}\,\boldsymbol{\partial}\right)$, implies
 \begin{eqnarray}
 \label{enmomcons}
    \int_V\mathrm{d}^3x\left(\frac{1}{c}\,\partial_0\,t^{00}+\sqrt{\frac{m_R}{m_I}}\,\partial_i\,t^{i0}\right)=0~,
 \end{eqnarray}
 where $V$ is the far region volume. The energy within this volume is by definition \cite{Maggiore:2007ulw}
 \begin{eqnarray}
     E_V=\int_V\mathrm{d}^3x\,t^{00}~.
 \end{eqnarray}
Using the above, we can rewrite Eq. (\ref{enmomcons}) as
\begin{eqnarray}
\label{entimemod}
    \frac{1}{c}\frac{\mathrm{d}E_V}{\mathrm{d}t}=\int_V\mathrm{d}^3x\,\frac{1}{c}\,\partial_0\,t^{00}&=&-\sqrt{\frac{m_R}{m_I}}\int_V\mathrm{d}^3x\,\partial_i\,t^{i0} \nonumber \\
    &=&-\sqrt{\frac{m_R}{m_I}}\int_S\mathrm{d}A\,n_i\,t^{i0} \nonumber \\
    &=&-\sqrt{\frac{m_R}{m_I}}\int_S\mathrm{d}A\,t^{r0}~,
\end{eqnarray}
where Gauss' theorem was used in the second line to write the expression as an integral over a far region surface $S$, and the normal vector to a spherical surface $\hat{\mathbf{n}}=\hat{\mathbf{r}}$ was used in the third line. The QEP modified expression for the far region energy-momentum tensor in the linearized theory follows from Eq. (\ref{qepeinstein}) and reads as
\begin{eqnarray}
    t^{\mu\nu}=\frac{m_R^2}{m_Gm_I}\,\frac{c^4}{32\pi G}\,\frac{m_I}{m_R}\,\langle\partial^\mu h_{\alpha\beta}\,\partial^{\nu}h^{\alpha\beta}\rangle~,
\end{eqnarray}
where the additional correction $m_I/m_R$ comes from the fact that $u^\mu u_\mu=m_R/m_I$, while $t^{\mu\nu}\propto u^\mu u^{\nu}$. We use the above to express the $t^{r0}$ from Eq. (\ref{entimemod}) as
\begin{eqnarray}
\label{apptensor}
    t^{r0}=\frac{m_R}{m_G}\,\frac{c^4}{32\pi G}\,\left\langle\sqrt{\frac{m_R}{m_I}}\,\partial_rh_{ij}^{TT}\,\frac{1}{c}\,\partial_0h_{ij}^{TT}\right\rangle~,
\end{eqnarray}
where $h_{\alpha\beta}$ is cast in the TT gauge, and has the general form of
\begin{eqnarray}
    h_{ij}^{TT}(t,r)=\frac{1}{r}\,f_{ij}\!\left(t-\sqrt{\frac{m_I}{m_R}}\frac{r}{c}\right)~,
\end{eqnarray}
where $f_{ij}(t_{ret})$ are functions of the retarded time, and their exact form is not relevant in this discussion. The above is used to find a relationship between the derivatives of $h_{ij}^{TT}$ in Eq. (\ref{apptensor}), which turn out as
\begin{eqnarray}
    \sqrt{\frac{m_R}{m_I}}\,\partial_rh_{ij}^{TT}=-\frac{1}{c}\,\partial_0h_{ij}^{TT}+\mathcal{O}\left(\frac{1}{r^2}\right)~,
\end{eqnarray}
from where it can be seen that $t^{r0}=t^{00}$ for large $r$. Therefore, we can rewrite Eq. (\ref{entimemod}) as
\begin{eqnarray}
\label{appeqenspec}
    \frac{\mathrm{d}E_V}{\mathrm{d}t}=-\sqrt{\frac{m_R}{m_I}}\,c\int_S \mathrm{d}A\,t^{00}\,\,\,\,\implies\,\,\,\,\frac{\mathrm{d}^2E}{\mathrm{d}A\,\mathrm{d}t}=\sqrt{\frac{m_R}{m_I}}\,c\,t^{00}=\frac{m_R^{3/2}}{m_Gm_I^{1/2}}\,\frac{c^3}{32\pi G}\,\langle\dot{h}_{ij}^{TT}\,\dot{h}_{ij}^{TT}\rangle~,
\end{eqnarray}
where $E$ is the energy carried away by the outward propagating GWs as opposed to the energy $E_V$, contained within a volume $V$, which results in the positive sign in the right-hand side equation. Since $\mathrm{d}A=r^2\mathrm{d}\Omega$, and taking the integral over time, the above returns the radiated energy in a solid angle $\mathrm{d}\Omega$ as
\begin{eqnarray}
    \frac{\mathrm{d}E}{\mathrm{d}\Omega}=\frac{m_R^{3/2}}{m_Gm_I^{1/2}}\,\frac{c^3r^2}{32\pi G}\int_{-\infty}^\infty\mathrm{d}t\,\dot{h}_{ij}^{TT}\,\dot{h}_{ij}^{TT}~,
\end{eqnarray}
where the fact that the average $\langle\cdots\rangle$ is a time average, makes the integral over time on Eq. (\ref{appeqenspec}), an integral over a constant. Therefore, the time integral returns the time interval over which the average is normalized. This cancels with the normalization time interval of the average and the remaining integral is the time integral, which was used to define $\langle\cdots\rangle$. Note that the above result is exactly the first line of Eq. (\ref{genenmod}).

 \section{Choice of the ground state mass}
 \label{proofpendix}

 Any violation of the EEP will be observable in terms of the ratio of different masses $m_\alpha/m_{\alpha'}$, where $\alpha,\alpha'=I,G,R$ and $\alpha\neq\alpha'$. After promoting $m_\alpha$ to a quantum operator $m_\alpha\longrightarrow\hat{M}_\alpha=m_\alpha\hat{\mathbb{1}}+\hat{H}_{int,\alpha}/c^2$, and after some algebraic manipulation, we obtain
 \begin{eqnarray}
     {\hat{M}_\alpha}\,{\hat{M}_{\alpha'}^{-1}}\approx\frac{m_\alpha}{m_{\alpha'}}\left(\hat{\mathbb{1}}+\frac{\hat{H}_{int,\alpha}}{m_\alpha c^2}-\frac{\hat{H}_{int,\alpha'}}{m_{\alpha'} c^2}\right)\equiv\hat{\mathbb{1}}+\hat{\delta}~,
 \end{eqnarray}
 where we consider ${\langle \hat{H}_{int,\alpha}\rangle}/{m_\alpha c^2}\ll1$, and $\hat{\delta}$ is the deviation of the ratio from unity operator. In the above, terms of order $\mathcal{O}(\hat{H}_{int,\alpha}^2)$ are neglected. By solving for $\hat{\delta}$, we obtain
 \begin{eqnarray}
 \label{qepdel}
     \hat{\delta}=\frac{m_\alpha}{m_{\alpha'}}\hat{\mathbb{1}}-\hat{\mathbb{1}}+\frac{m_\alpha}{m_{\alpha'}}\left(\frac{\hat{H}_{int,\alpha}}{m_\alpha c^2}-\frac{\hat{H}_{int,\alpha'}}{m_{\alpha'} c^2}\right)~.
 \end{eqnarray}
 Here, we can choose that masses $m_\alpha$ differ from each other as
 \begin{eqnarray}
     m_\alpha=m+\Delta_\alpha~,
 \end{eqnarray}
 where $m$ is the common ground state for all different masses and $\Delta_\alpha$ the classical deviation from $m$ (if any) for a given $\alpha$. Note that $\hat{H}_{int,\alpha}$ and $\Delta_\alpha$ can in principle have different origins. By plugging the above $m_\alpha$ in Eq. (\ref{qepdel}), and neglecting all terms $\mathcal{O}(\hat{H}_{int,\alpha}\,\Delta_{\alpha'})$, we obtain
 \begin{eqnarray}
     \hat{\delta}\approx\frac{\hat{\tilde{H}}_{int,\alpha}}{m c^2}-\frac{\hat{\tilde{H}}_{int,\alpha'}}{m c^2}~,
 \end{eqnarray}
 where $\hat{\tilde{H}}_{int,\alpha}=\hat{H}_{int,\alpha}+\Delta_\alpha c^2\hat{\mathbb{1}}$. This means that we can always choose a set of operators $\hat{H}_{int,\alpha}$, which include all sources of deviations, such that the different mass operators $\hat{M}_\alpha$ have the exact same ground state $m$.

 \section{EEP tests}
 \label{qeptests}

To test the EEP, we need to test the individual aspects of LLI, WEP and LPI. Note that the operator equality $\hat{M}_R=\hat{M}_I$ represents the quantum version of LLI, $\hat{M}_G=\hat{M}_I$ represents the quantum version of WEP and $\hat{M}_G=\hat{M}_R$ represents the quantum version of LPI. For example, the predicted radiated energy is given by Eq. (\ref{predraden}), where the QEP modification appears as $m_R/m_I$. To test the LLI in the context of QEP, we need to promote these masses in the given ratio to quantum operators
 \begin{eqnarray}
     \frac{m_R}{m_I}\longrightarrow\hat{M}_R\hat{M}_I^{-1}~.
 \end{eqnarray}
 Following Appendix \ref{proofpendix} and evaluating the expectation values of operators $\hat{H}_{int,\alpha}$, i.e., $\langle\hat{H}_{int,\alpha}\rangle\equiv E_\alpha$ since this is what can be in principle directly measured, the deviation due to a violation of LLI reads as
 \begin{eqnarray}
 \label{llitest}
     \delta_{LLI}\approx\frac{1}{mc^2}(E_R-E_I)~.
 \end{eqnarray}
 For example, the predicted radiated energy from Eq. (\ref{predraden}) then reads as
 \begin{eqnarray}
     \Delta E_{rad}=\left(1-\sqrt{\frac{8}{9}}\right)\mu c^2 \left(1+\frac{1}{mc^2}(E_R-E_I)\right)~.
 \end{eqnarray}

 The same procedure can be used to predict deviations due to violations of WEP and LPI, where the deviation due to a violation of WEP can be written as
 \begin{eqnarray}
 \label{weptest}
     \delta_{WEP}\approx\frac{1}{mc^2}(E_G-E_I)~,
 \end{eqnarray}
 and the deviation due to a violation of LPI as
 \begin{eqnarray}
 \label{lpitest}
     \delta_{LPI}\approx\frac{1}{mc^2}(E_G-E_R)~.
 \end{eqnarray}

 In case where the modifications do not appear as ratios of two masses, as seen above, we obtain a modification on the form 
 \begin{eqnarray}
m_G^k\,m_R^l\,m_I^n\longrightarrow\hat{M}_G^k\hat{M}_R^l\hat{M}_I^n~,
 \end{eqnarray}
 where $k,l,n\in \mathbb{Q}$. By following Appendix \ref{proofpendix} evaluating the expectation values of $\hat{H}_{int,\alpha}$, and considering $E_\alpha/mc^2\ll1$, we obtain
 \begin{eqnarray}
     \langle\hat{M}_G^k\hat{M}_R^l\hat{M}_I^n\rangle\approx m^{k+l+n}\left(1+k\frac{E_G}{mc^2}+l\frac{E_R}{mc^2}+n\frac{E_I}{mc^2}\right)~.
 \end{eqnarray}
 A general QEP violation deviation then reads
 \begin{eqnarray}
     \delta_{QEP}=\frac{1}{mc^2}(kE_G+lE_R+nE_I)~.
 \end{eqnarray}
 If there are at least two measurements with this form of deviation but with different $k,l,n$, we can construct specific tests for WEP, LLI and LPI, as seen in Eqs. (\ref{llitest}), (\ref{weptest}) and (\ref{lpitest}), by solving a system of two equations for a specific test.

 \section{Bounds on frequency and signal measurements}
 \label{boundpendix}

 The bounds seen in table \ref{tab:bounds} are obtained by taking the measurement uncertainties on the chirp masses and on the event distance, obtained from Refs. \cite{Abbot:2020a,Abbot:2020b,Abbot:2020}. From Eqs. (\ref{hthere+}), (\ref{htherex}) and (\ref{modinspfreq}), we can see that the amplitude depends on the chirp mass $M_c$ and distance $r$, while the time dependent frequency depends on only the chirp mass $M_c$. The propagation of error method is then used to obtain these bounds. The uncertainty on the frequency measurement is obtained as
 \begin{eqnarray}
     \sigma_f=\pm\left(\frac{\partial f_{GW}}{\partial M_c}\right)\sigma_{M_c}^{U,L}~,
 \end{eqnarray}
 where $\sigma_{M_c}^{U,L}$ are the upper and lower bounds on $M_c$. The above determines $\mathrm{UB}_f=\sigma_f/f_{GW}$ for the $+$ solution with $\sigma_{M_c}^{U}$, and $\mathrm{LB}_f=\sigma_f/f_{GW}$ for the $-$ solution with $\sigma_{M_c}^{L}$. In a similar way, the uncertainty on the amplitude measurement is obtained as
 \begin{eqnarray}
     \sigma_A=\pm\sqrt{\left(\frac{\partial A}{\partial r}\right)^2\left(\sigma_r^{U,L}\right)^{2}+\left(\frac{\partial A}{\partial M_c}\right)^2\left(\sigma_{M_c}^{U,L}\right)^{2}}~,
 \end{eqnarray}
 where $\sigma_{r}^{U,L}$ are the upper and lower bounds on $r$ and $A=-\left(\frac{GM_c}{c^2}\right)^{5/4}\left(\frac{5}{c\tau}\right)^{1/4}\frac{1}{r}$ the amplitude of the unmodified GW wavefront, as seen in Eqs. (\ref{hthere+}) and (\ref{htherex}). The above determines $\mathrm{UB}_A=\sigma_A/A$ for the $+$ solution with $\sigma_{r}^{U}$ and $\sigma_{M_c}^{U}$, and $\mathrm{LB}_A=\sigma_A/A$ for the $-$ solution with $\sigma_{r}^{L}$ and $\sigma_{M_c}^{L}$.

 %To obtain the modifications to the form factors, one follows the derivation in Ref. \cite{Maggiore:2007ulw}. The derivation starts by considering the geodesic equation of a deviation $\xi^\mu$ between two geodesics $x^\mu$ and $x^\mu+\xi^\mu$, as
 %\begin{eqnarray}
 %\label{devgeod}
 %    \frac{\mathrm{d}^2\xi^\mu}{\mathrm{d}\tau^2}+2\,\Gamma_{\nu\rho}^\mu(x)\frac{\mathrm{d}x^\nu}{\mathrm{d}\tau}\frac{\mathrm{d}\xi^\rho}{\mathrm{d}\tau}+\xi^\sigma\partial_\sigma\Gamma_{\nu\rho}^\mu(x)\frac{\mathrm{d}x^\nu}{\mathrm{d}\tau}\frac{\mathrm{d}x^\rho}{\mathrm{d}\tau}=0~.
 %\end{eqnarray}
 %One can then choose a coordinate system in which the Christoffel symbols vanish at the expansion point $\Gamma_{\nu\rho}^\mu=0$, and because the mirror movements are non-relativistic, $\frac{\mathrm{d}x^i}{\mathrm{d}\tau}\ll\frac{\mathrm{d}x^0}{\mathrm{d}\tau}$. Therefore, one can write Eq. (\ref{devgeod}) as
 %\begin{eqnarray}
 %    \frac{\mathrm{d}^2\xi^i}{\mathrm{d}\tau^2}+\xi^\sigma\partial_\sigma\Gamma_{00}^i\left(\frac{\mathrm{d}x^0}{\mathrm{d}\tau}\right)^2=0~.
 %\end{eqnarray}
 %As shown in Ref. \cite{Das:2023cfu}, one can write the last bracket as
 %\begin{eqnarray}
 %    \left(\frac{\mathrm{d}x^0}{\mathrm{d}\tau}\right)^2=\frac{m_R}{m_I}c^2~,
 %\end{eqnarray}


\begin{thebibliography}{99}

 %\cite{Zeilinger:1999zz}
\bibitem{Zeilinger:1999zz}
A.~Zeilinger,
%``Experiment and the foundations of quantum physics,''
Rev. Mod. Phys. \textbf{71}, S288-S297 (1999).
%doi:10.1103/RevModPhys.71.S288
%71 citations counted in INSPIRE as of 12 Dec 2023

 \bibitem{CMW}
C.~M.~Will,
The Confrontation between General Relativity and Experiment,
Living Rev. Rel. \textbf{17}, 4 (2014).
%doi:10.12942/lrr-2014-4
%[arXiv:1403.7377 [gr-qc]].

\bibitem{qg1}
B.~Zwiebach. {A First Course in String Theory}, Cambridge University Press: Cambridge, UK (2004)


%\cite{Smolin:2004sx}
\bibitem{Smolin:2004sx}
L.~Smolin,
%``An Invitation to loop quantum gravity,''
doi:10.1142/9789812702340\_0078
[arXiv:hep-th/0408048 [hep-th]].
%198 citations counted in INSPIRE as of 06 Feb 2024

\bibitem{Einstein}
A.~Einstein,
{The Meaning of Relativity},
Princeton University Press: Aberdeen, UK (1922)

%\cite{Weinberg:1972kfs}
\bibitem{Weinberg:1972kfs}
S.~Weinberg,
{Gravitation and Cosmology: Principles and Applications of the General Theory of Relativity}, John Wiley and Sons: New York, USA (1972)
%210 citations counted in INSPIRE as of 12 May 2022

%\cite{Zych:2015fka}
\bibitem{Zych:2015fka}
M.~Zych and \v{C}.~Brukner,
Quantum formulation of the Einstein Equivalence Principle,
Nature Phys. \textbf{14}, no.10, 1027-1031 (2018).
%doi:10.1038/s41567-018-0197-6
%[arXiv:1502.00971 [gr-qc]].
%72 citations counted in INSPIRE as of 10 Feb 2022

%\cite{Das:2023cfu}
\bibitem{Das:2023cfu}
S.~Das, M.~Fridman and G.~Lambiase,
%``General formalism of the quantum equivalence principle,''
Commun. Phys. \textbf{6}, no.1, 198 (2023)
%doi:10.1038/s42005-023-01306-w
[arXiv:2307.09632 [gr-qc]].
%0 citations counted in INSPIRE as of 17 Jan 2024

%\cite{Feng:2016tyt}
\bibitem{Feng:2016tyt}
Z.~W.~Feng, S.~Z.~Yang, H.~L.~Li and X.~T.~Zu,
%``Constraining the generalized uncertainty principle with the gravitational wave event GW150914,''
Phys. Lett. B \textbf{768}, 81-85 (2017)
%doi:10.1016/j.physletb.2017.02.043
[arXiv:1610.08549 [hep-ph]].
%85 citations counted in INSPIRE as of 05 Feb 2024

%\cite{Katsuragawa:2019uto}
\bibitem{Katsuragawa:2019uto}
T.~Katsuragawa, T.~Nakamura, T.~Ikeda and S.~Capozziello,
%``Gravitational Waves in $F(R)$ Gravity: Scalar Waves and the Chameleon Mechanism,''
Phys. Rev. D \textbf{99}, no.12, 124050 (2019)
%doi:10.1103/PhysRevD.99.124050
[arXiv:1902.02494 [gr-qc]].
%40 citations counted in INSPIRE as of 05 Feb 2024

%\cite{Bhattacharyya:2020ooz}
\bibitem{Bhattacharyya:2020ooz}
S.~Bhattacharyya, S.~Gangopadhyay and A.~Saha,
%``Generalized uncertainty principle in resonant detectors of gravitational waves,''
Class. Quant. Grav. \textbf{37}, no.19, 195006 (2020)
%doi:10.1088/1361-6382/abac45
[arXiv:2005.09454 [gr-qc]].
%15 citations counted in INSPIRE as of 05 Feb 2024

\bibitem{Gogoi:2020}
D.~J.~Gogoi and U.~Dev Goswami,
%``A new f(R) gravity model and properties of gravitational waves in it,''
Eur. Phys. J. C \textbf{80}, no.12, 1101 (2020)
doi:10.1140/epjc/s10052-020-08684-3.

%\cite{Das:2021lrb}
\bibitem{Das:2021lrb}
A.~Das, S.~Das, N.~R.~Mansour and E.~C.~Vagenas,
%``Bounds on GUP parameters from GW150914 and GW190521,''
Phys. Lett. B \textbf{819}, 136429 (2021)
%doi:10.1016/j.physletb.2021.136429
[arXiv:2101.03746 [gr-qc]].
%23 citations counted in INSPIRE as of 05 Feb 2024

%\cite{Moussa:2021qlz}
\bibitem{Moussa:2021qlz}
M.~Moussa, H.~Shababi and A.~Farag Ali,
%``Generalized uncertainty principle and stochastic gravitational wave background spectrum,''
Phys. Lett. B \textbf{814}, 136071 (2021)
%doi:10.1016/j.physletb.2021.136071
[arXiv:2101.04747 [gr-qc]].
%7 citations counted in INSPIRE as of 05 Feb 2024

%\cite{Kalita:2021zjg}
\bibitem{Kalita:2021zjg}
S.~Kalita and B.~Mukhopadhyay,
%``Gravitational wave in f(R) gravity: possible signature of sub- and super-Chandrasekhar limiting mass white dwarfs,''
Astrophys. J. \textbf{909}, no.1, 65 (2021)
%doi:10.3847/1538-4357/abddb8
[arXiv:2101.07278 [astro-ph.HE]].
%14 citations counted in INSPIRE as of 05 Feb 2024

\bibitem{Lambiase:2020vul}
G. Lambiase, M. Sakellariadou, and A. Stabile,
%    title = "{Constraints on extended gravity models through gravitational wave emission}",
%    eprint = "2012.00114",
%    archivePrefix = "arXiv",
%    primaryClass = "gr-qc",
%    doi = "10.1088/1475-7516/2021/03/014",
JCAP03 (2021) 014,

\bibitem{Tino:2020nla}
%Precision Gravity Tests and the Einstein Equivalence Principle
G.M. Tino, L. Cacciapuoti, S. Capozziello, G. Lambiase, F. Sorrentino, Prog.Part.Nucl.Phys. 112 (2020) 103772.

%\cite{Das:2022hjp}
\bibitem{Das:2022hjp}
S.~Das, S.~Shankaranarayanan and V.~Todorinov,
%``Quantum gravitational signatures in next-generation gravitational wave detectors,''
Phys. Lett. B \textbf{835}, 137511 (2022)
%doi:10.1016/j.physletb.2022.137511
[arXiv:2208.11095 [gr-qc]].
%3 citations counted in INSPIRE as of 05 Feb 2024

 %\cite{Maggiore:2007ulw}
\bibitem{Maggiore:2007ulw}
M.~Maggiore,
``Gravitational Waves. Vol. 1: Theory and Experiments,''
Oxford University Press, 2007,
ISBN 978-0-19-171766-6, 978-0-19-852074-0
%doi:10.1093/acprof:oso/9780198570745.001.0001
%171 citations counted in INSPIRE as of 20 Jul 2023

\bibitem{Abbot:2020a}
R. Abbott et al, Phys. Rev. Lett. \textbf{119}, 161101 (2017)


\bibitem{Abbot:2020b}
R. Abbott et al, Phys. Rev. Lett. \textbf{125}, 101102 (2020)

\bibitem{Abbot:2020}
R. Abbott et al, ApJL \textbf{896}, L44 (2020)

\bibitem{refId0}
{M. H. P. M. van Putten} and {M. Della Valle},
{``Central engine of GRB170817A: Neutron star versus Kerr black hole based on multimessenger calorimetry and event timing''},
"10.1051/0004-6361/202142974",
%"https://doi.org/10.1051/0004-6361/202142974",
A\&A, \textbf{669}, A36 (2023)

     
 \end{thebibliography}
\end{document}